\documentclass[letterpaper,twocolumn,amsmath,showpacs,amssymb,aps,nofootinbib]{revtex4-1}
\usepackage[T1]{fontenc}
\usepackage[utf8]{inputenc}
\usepackage{amsfonts}
\usepackage{amssymb}
\usepackage{lmodern}
\usepackage{graphicx}
\usepackage{mathtools}
\usepackage{verbatim}
\usepackage[dvipsnames]{xcolor}
\usepackage{geometry}
\usepackage{amsmath,amssymb,amsfonts,dcolumn,color,graphicx,graphics,latexsym,placeins,epsfig,tikz}
\usetikzlibrary{arrows,shapes}
\usepackage{subfigure,rotating,bm,mathrsfs}

\newcommand{\RN}[1]{
	\textup{\uppercase\expandafter{\romannumeral#1}}
}
\newcommand{\be}{\begin{equation}}
	\newcommand{\ee}{\end{equation}}
\newcommand{\ba}{\begin{eqnarray}}
	\newcommand{\ea}{\end{eqnarray}}
\newcommand{\bs}{\begin{split}}
	\newcommand{\es}{\end{split}}

\usepackage[pagebackref=false, colorlinks=true]{hyperref}
\definecolor{redish}{rgb}{0.7,0.2,0.0}  
\definecolor{bluish}{rgb}{0.2,0.5,0.8}
\hypersetup{linkcolor=red,          
	citecolor=SeaGreen,        
	filecolor=magenta,      
	urlcolor=blue}          

\begin{document}
\title{\Large Complexity in two-point measurement schemes}
\author{Ankit Gill}\email{ankitgill20@iitk.ac.in}
\affiliation{
Department of Physics, Indian Institute of Technology Kanpur, \\ Kanpur 208016, India}
\author{Kunal Pal}\email{kunalpal@iitk.ac.in}
\affiliation{
Department of Physics, Indian Institute of Technology Kanpur, \\ Kanpur 208016, India}
\author{Kuntal Pal}\email{kuntal@iitk.ac.in}
\affiliation{
Department of Physics, Indian Institute of Technology Kanpur, \\ Kanpur 208016, India}
\author{Tapobrata Sarkar}\email{tapo@iitk.ac.in}
\affiliation{
Department of Physics, Indian Institute of Technology Kanpur, \\ Kanpur 208016, India}

\begin{abstract}
We show that the characteristic function of the probability distribution associated with the change of an observable in a two-point measurement 
protocol with a perturbation can be written as an auto-correlation function between an initial state and a certain unitary evolved state by an effective unitary operator. 
Using this identification, we 
probe how the evolved state spreads in the corresponding conjugate space, by defining a notion of the complexity of the spread of this evolved state. For a 
sudden quench scenario, where the parameters of an initial Hamiltonian (taken as the observable measured in the two-point measurement protocol) 
are suddenly changed to a new set of values, we first obtain the corresponding Krylov basis vectors and the associated Lanczos 
coefficients for an initial pure state, and obtain the spread complexity. Interestingly, we find that in such a protocol, the Lanczos coefficients can be 
	related to various cost functions used in the geometric formulation of circuit complexity, for example the one used to define Fubini-Study complexity.
We illustrate the evolution of spread complexity both analytically, by using  Lie algebraic 
techniques, and by performing numerical computations. This is done for cases when the Hamiltonian before and after the quench are taken as different combinations 
of chaotic and integrable spin chains. We show that the complexity saturates for large values of the parameter only when the pre-quench Hamiltonian is chaotic. 
Further, in these examples we also discuss the important  role played by the initial state which is  determined by the time-evolved perturbation operator.

\end{abstract}
\maketitle

\section{Introduction}

The dynamics of an isolated quantum system, which is taken out-of equilibrium, is a topic of great recent interest. This is
partly due to the fact that from the  experimental point of view, modern ultra-cold atoms provide an excellent approximation of an 
isolated quantum system \cite{Polkovnikovrivew, Bloch}. Among the most important questions raised to explain 
these path-breaking experiments, one of the most fundamental is the possible mechanism behind the  emergence of classical
thermodynamics from underlying quantum statistical mechanics. 
To this end, it is of primary importance to note that the commonly used concept of work associated with a process in classical thermodynamics is no longer an observable 
in one of the canonical ways of defining the analogous quantity in quantum systems, namely, in a so-called two point measurement (TPM) scheme
\cite{Kurchan, Tasaki, Mukamel, Talkner1}. The reason is that the definition 
of work $\mathcal{W}$ associated with a quantum process (as measured through a TPM scheme), involves two projective energy measurements of the system, one at the initial time before the process, and another after the unitary evolution by a second Hamiltonian after the 
time we want to measure the system. As a result, no Hermitian operator can be associated with $\mathcal{W}$, and hence it is not an observable \cite{Talkner1}. Actually, one needs to perform the above measurements on infinitely many realisations of the same system, and the work is described by a probability distribution $P(\mathcal{W})$. 
The fluctuation of work in a generic thermal quantum system then comes from both the thermal and quantum mechanical fluctuations. 

An important quantity that elucidates many universal features associated with the above notion of work statistics is the Fourier transform of the 
work probability distribution function, i.e., the characteristic function (CF) of work distribution
\begin{equation}\label{CF}
G(u)=\int P(\mathcal{W}) e^{-i u \mathcal{W}} d \mathcal{W}~.
\end{equation}
Here, the auxiliary variable $u$ is the conjugate of the work done on the system. Importantly, as shown in \cite{Talkner1} (see also \cite{Silva}), the  above CF can be interpreted
as a correlation function corresponding to the $u$-evolution.  For example, in a sudden quench protocol, where one changes the 
parameters of an initial Hamiltonian suddenly at an instant of time to a new set of values,  the CF is just the Loschmidt amplitude, i.e., the overlap between the initial state and a  $u$-evolved state\cite{Silva}.  In a related but somewhat different context, the work of \cite{Campisi} showed how the CF of the work distribution (WD) in the TPM protocol with a perturbation operator inserted instantly, can be written down as an out-of-time ordered correlator (OTOC) \cite{Larkin, Maldacena} between certain operators. As a result, it was also possible to draw a connection between the dynamics of information scrambling, as quantified by the OTOC, and thermodynamically fluctuating quantities like work in such a protocol. 

Another important  relation connecting the quantum thermodynamic quantities like WD and the Loschchmidt echo (LE) was established in \cite{Chenu1}, where it was shown that for an isolated quantum system in a generic mixed state, the CF of work under a quantum quench is related to the LE amplitude in a larger Hilbert space of an auxiliary system, representing the purification of the initial density matrix, and that the quench can be thought of as acting on a single copy of this auxiliary system. In this context, it can be noted that the LE, which measures the overlap of an initial state with a state that has undergone a forward evolution by the system Hamiltonian, and a subsequent backward time evolution by a slightly perturbed Hamiltonian, is a widely used measure in quantum chaos literature \cite{Peres, Jalabert, Goussev, Gorin}. 

This set of results points towards an underlying connection between the fluctuations of thermodynamic quantities associated 
with the non-equilibrium dynamics of an isolated quantum system and the chaotic or integrable nature of the Hamiltonian governing the unitary dynamics itself. 
As the CF contains the dynamics of an auxiliary system where the role of time is played by the scalar parameter (denoted in Eq. (\ref{CF}) as $u$) which,
in general, is the  Fourier conjugate variable of the eigenvalue of  the operator which is measured in the TPM scheme, this also encodes the full information of work PDF. In particular, this connection was firmly established in  \cite{Chenu1}, where it was shown that for certain type of quench protocols with chaotic Hamiltonians, the features of scrambling of information as encoded in the LE, which shows the characteristic dip, ramp, and plateau structure, can also be accessed by the corresponding CF, and hence the associated WD.
This is the link we pursue further in the present paper by using the tools of quantum complexity theory, particularly the Krylov complexity (KC),  a new measure of the `complexity' of a time-evolved  operator that has been used extensively in recent times to probe the dynamics quantum chaotic systems and the physics of information scrambling.

The KC, a new addition to the dictionary of the complexity of quantum systems, was originally introduced in the context of measuring operator growth in quantum-many-body systems in \cite{Parker:2018yvk}. After the original work, the central result of which introduced a hypothesis about the growth of the so-called Lanczos coefficients (LCs), KC has become a very fruitful measure to study various aspects of quantum systems, both in and out-of-equilibrium. For a partial set of works, see \cite{Barbon:2019wsy, Bhattacharjee:2022vlt} 
for the use of KC in operator growth, \cite{Dymarsky:2019elm, Dymarsky:2021bjq, Kundu:2023hbk} for works in CFTs, 
\cite{Bhattacharya:2022gbz, Bhattacharjee:2022lzy, Bhattacharya:2023zqt} for works on open systems, \cite{Bhattacharyya:2023dhp} for 
KC in Bosonic systems, \cite{Kim:2021okd} as a tool of probing 
delocalization properties in integrable quantum systems, \cite{Hashimoto:2023swv, Camargo:2023eev} focuses on billiard systems. 
Important steps have also been taken to understand features of KC in  QFTs \cite{ Camargo:2022rnt, Avdoshkin:2022xuw}. Various other features and uses of KC were also explored in \cite{Caputa:2021ori}-\cite{Fan:2023ohh}.\footnote{There are a large number of works on various aspects of KC and operator growth, so the list of references mentioned above is incomplete. For a  more complete account  of various avenues explored the reader can look at the citations of the papers mentioned here.}

A related but somewhat different concept of the so-called Krylov state complexity or spread complexity (SC) was first introduced in \cite{Balasubramanian:2022tpr}, where it was proved that a certain cost associated with the spread of a quantum state under a Hamiltonian evolution with respect to a fixed set of basis vector on the Hilbert space is minimised only if the basis vector is taken as the Krylov basis
generated by the Hamiltonian. The Krylov basis is a set of orthonormal vectors on the Hilbert space that can be constructed using the Lanczos algorithm \cite{Viswanath, Lanczos:1950zz}, and for a Hamiltonian evolution, this algorithm generally gives two sets of LCs. After the original work of \cite{Balasubramanian:2022tpr}, SC in 
quantum systems has been explored in various papers that 
include quantum phase transition \cite{Caputa2, Caputa:2022yju, spread1}, work statistics in quantum quenches \cite{Pal:2023yik, Gautam:2023pny}, 
probing quantum scar states \cite{Nandy:2023brt}, systems described by random matrices \cite{Erdmenger:2023shk}, studying integrability to chaos transition \cite{Scialchi:2023bmw}, interacting quantum systems \cite{SCint} among others. 

One of our primary goals is to unify two different kinds of observables, namely, quantum information theoretical (such as the entanglement entropy, OTOC, complexity etc.,  see \cite{Igloi, Calabrese1, Calabrese2, Alaves, Camargo}) and quantum thermodynamical (such as the work performed, heat generated etc., see \cite{Barankov:2008qq, Polkovnikov, Silva}),  
that are commonly employed to study evolution of a quantum  system subjected to quenches.
In our previous work \cite{Pal:2023yik}, we obtained one such relations, where we showed that the  LCs associated with the evolution
generated by the post-quench Hamiltonian can be obtained from the average, variance, and higher order cumulants of the distribution of the  work done on a system through a sudden quench, thereby providing interpretations of these coefficients in terms of experimentally observable quantities. 

In this work, we proceed with a similar motivation, and first discuss a connection between the two concepts mentioned in the previous paragraphs, namely CF associated with probability distribution of an observable in a TPM scheme and the complexity of the spread of a certain unitary evolved state, by noting that the CF of a TPM scheme (with a perturbation operator $W$ introduced between a forward and backward
evolution) can be equivalently  viewed as an auto-correlation function (ACF) corresponding to the evolution generated by the hermitian operator
(denoted as $O$ in subsequent analysis) which is measured under a TPM scheme, see section 
\ref{otocacf}.  This naturally leads us to define a set of Krylov bases generated 
by the observable $O$, and the concept of SC associated with the $u$-evolved state in section \ref{SCTPM}.
 This will help us to probe the unitary evolution 
in the space that is Fourier conjugate to the observable $O$ -- the change of which is measured in a TPM protocol -- by using the tools of Krylov state complexity.  The fact that the full set of LCs, and consequently the full Krylov basis sets, can be extracted from the knowledge of the ACF, we are able to study the behaviour of SC both analytically and numerically, in section \ref{algebra} and \ref{spinchain}, respectively. 
In particular, we will show how the integrable or chaotic nature of the operator $O$ (which can be taken as the Hamiltonian of some quantum system, $H$) differentiates the spreading of information in the conjugate $u$-space. 
 
In a similar spirit, we  establish relations between the fidelity OTOC (FOTOC), the survival probability of the $u$-evolved state, and the corresponding  LCs, and 
show that in the sudden quench protocol, when $O$ is the Hamiltonian $H_0$, the LC $b_1$ is the Fisher information
of the time-evolved state. Furthermore, in section \ref{iprb1}, we show that  the nature of the inverse participation ratio (IPR) 
of the time-evolved state 
is directly related to the nature of the LCs of the $u$-evolved state, and verify this connection using an analytical example
with the Hamiltonian taken as an element of the $su(1,1)$ Lie algebra. Finally, in section \ref{conclusion}, we discuss possible
 importance of our results from the 
point of view experiments that are used to measure the WD, and the subsequent possibility of relating SC with experimentally 
measurable quantities.

\section{Out-of-time-order correlators  from two-point measurement protocols}
The OTOC was first  introduced in \cite{Larkin} to study the instability of electron trajectories in superconductors. 
For two generic hermitian (or unitary operators) $V$ and $W$,  it is defined as
\begin{equation}\label{sqcomm}
	C_{V,W}(t)=\big<[W_t,V]^{\dagger}[W_t,V]\big>~,
\end{equation}
where $W_t=e^{- i H t} W e^{ i H t}$ represents time evolved form of the initial operator $W=W(t=0)$ under the Hamiltonian $H$.  
After the work of \cite{Maldacena}, this quantity
has been used extensively to study quantum chaos. Expanding the expression for the correlator $C_{V,W}(t)$, we see that it contains
terms where the operators are ordered in out-of-time  fashion (in contrast to the usual time-ordered correlators). Among such terms, in this 
paper, our interest will be a term of the form 
\begin{equation}\label{OTOC}
	F_{V,W}(t)=\big<W_t^\dagger V^{\dagger}W_t V\big>~,
\end{equation}
which, as we shall see, can be re-casted as an ACF. Since,
$\text{Re}[F_{V,W}(t)]=1-\big<[V^\dagger, W_t^\dagger][W_t,V]\big>/2$, we see that this quantity actually measures the amount by 
which the operators $W_t$ and $V$ fail to commute at a later time $t$ under the evolution by the Hamiltonian $H$, provided that the operators $W$ and $V$ commute at the initial time $t=0$.

In quantum systems, the OTOC can be used as one of the diagnostic tools of quantum chaos at the level of dynamics. For systems that have a well 
defined semiclassical limit, or systems that have a large number of local degrees of freedom, one can characterise quantum chaos from the
short-time exponential growth of OTOC \cite{Maldacena}. 
Physically, OTOC measures how quantum information, which was initially in some local subsystem, becomes delocalised in the entire system.
This `spread' of local information to the entire system is usually called the scrambling of quantum information.   
One can quantify this spreading by using the growth of a local operator under a Hamiltonian evolution. Mathematically, one way to do
this is to use the squared commutator in Eq. \eqref{sqcomm}, where, as we have mentioned above, two initially commutating operators  
will no longer commute due to the fact that the operator $W$ gets  `complicated' due to the  time evolution.  

Another way 
one can measure how an operator or a quantum state gets complicated under a Hamiltonian evolution is 
by counting the support of a time-evolved state
or an operator in terms of a specified  orthonormal basis, known as the Krylov basis. The resulting measure,  the KC or the SC 
discussed in the Introduction has gained wide attention recently due to the fact that it is another very useful tool for characterising  quantum chaos, since the corresponding 
sets of LCs as well as this measure of complexity show particular behaviour for quantum chaotic systems \cite{Parker:2018yvk,Balasubramanian:2022tpr}.  In this paper, we define
a special class of SC  from the OTOC, which quantifies how the initial  state of a quantum system that has been perturbed far from  equilibrium 
through a sudden quench, spreads under evolution generated by the initial Hamiltonian.

\subsection{Two-point measurement schemes}\label{TPM}

We start by briefly describing the protocol considered in this paper.
Due to the appearance of  out-of-time order operators discussed above,  it is difficult to measure correlators experimentally. 
Recently, using the well-known TPM scheme \cite{Kurchan,Tasaki},
in \cite{Campisi}, the authors have proposed an alternative method of measuring these
for a wide class of states. In such a TPM scheme, an observable $O$ is projectively measured before and 
after a non-equilibrium process (such as 
a quantum quench) is performed on a quantum system. 

In \cite{Campisi}, one such protocol was considered, and it contains the following
steps: (1) A quantum system  (with a  Hamiltonian denoted by $H_0$) is prepared in some state $\rho$ at $t=0$. 
(2) First projective measurement of an observable $O$ is performed, after which the system collapses to an eigenstate $|O_n\big>$ of the 
observable, and giving a result $O_n$, the $n$th eigenvalue
of the operator $O$. 
(3)  For $t>0$ the system is evolved with a Hamiltonian 
$H$ for a time $t=\tau$. 
(4) After this time, a unitary perturbation $W$ (known as the wing-flap operator) is applied to the system.
(5)  The system is evolved with $-H$ for a a time  $t=\tau$, and finally, 
(6) a second projective measurement of $O$ is performed which now
yields a value $O_m$, thereby collapsing the system to the eigenstate $|O_m\big>$ of $O$. For a schematic representation of this 
protocol, see fig. 1 of reference \cite{Campisi}.

Due to the presence of the perturbation $W$, the state of the system after the evolution with $-H$ (the backward evolution, step (5)
of the above protocol) is not the same as the initial state $|n\big>$; rather, it can be thought of as a linear combination 
of all the eigenstates of $O$. Therefore, the second projective measurement can select any one of these states, and hence
$|O_m\big>$ can be different from the initial state.

Since after the second  measurement the eigenvalue can take any of the possible values $O_m$, by repeating the above protocol 
a very large number of times, we can obtain a probability distribution function (PDF) for the change in  the value of the 
operator $O$ due to the perturbation $W$. The PDF for the change of the observable $\Delta O= O_m-O_n$ is given by
the expression 
\begin{equation}\label{pdfo}
P(\Delta O, \tau)=\sum_{n,m} P_{\tau}[O_m|O_n] p_n \delta (\Delta O -(O_m-O_n))~,
\end{equation}
where $p_n$  denotes the probability of getting an outcome $O_n$ after the first measurement of the observable $O$, and
$P_{\tau}[m|n]$ gives the probability of obtaining the result $O_m$ after the second measurement with the condition that the
first measurement yields a value $O_n$. For the TPM protocol described above, the expression for this quantity can be written as
\begin{align}\label{distribution}
P_{\tau}[O_m|O_n] & =\big|\big<O_m\big|e^{i\tau H} W e^{-i\tau H}\big|O_ n\big>\big|^2~\nonumber\\
& =\big|\big<O_m\big|W_\tau\big|O_n\big>\big|^2~,
\end{align}
where $W_\tau$ denotes the operator $W$ at time $\tau$ in the Heisenberg representation,
i.e., $ W_{\tau} = e^{- i H \tau} W e^{ i H \tau} $. Here we assume that the observable $O$ has non-degenerate spectrum and that the 
condition $[\rho, O]=0$ is satisfied. 

Next we need to consider the CF of the above PDF in Eq. \eqref{pdfo}, which is defined as  the Fourier transform
of $P(\Delta O, \tau)$,
\begin{equation}\label{CF1}
	G(u,\tau)=\int P(\Delta O, \tau) e^{-i u \Delta O} d (\Delta O)~.
\end{equation}
The auxiliary variable $u$, which appears as the conjugate of $\Delta O$, is sometimes called the second time of evolution
in the literature \cite{Silva, Chenu1}.

The significance of this quantity for our purposes can be understood as follows. Suppose we perform a sudden change in the 
parameters of a quantum system, i.e., the  system is subjected to a quench, so that the Hamiltonian of the system is changed 
from $H_0$ at $t=0$ to $H$ and the system is subsequently evolved with the new Hamiltonian.
If we measure the energy of the system before and after
such a change, i.e., the operator $O$ is the Hamiltonian of the system, 
then the change in the energy $\mathcal{W}=E_n^0-E_m$ of the system is the work done on the system due to this  quench.
Assuming that the system was prepared in an eigenstate $\big| 0\big>$ of the pre-quench Hamiltonian, the CF of the WD can be re-casted as  a correlation function of the form \cite{Talkner1}
\begin{equation}\label{CF2}
		G(t)=\langle 0\big|e^{i H_{0} t}e^{-i H t}\big|0\rangle~
		=e^{i E_{0} t} \langle \psi_0\big|\Psi(t)\rangle~,
\end{equation}
where $\big|\Psi(t)\rangle$ denotes the time evolved state after the quench, and $t$ denotes the 
time after the quench, and here it is conjugate to the work done $\mathcal{W}$. As was established in \cite{Silva}, this
is just the Loschmidt amplitude,\footnote{This is also the ACF between the time-evolved state and the initial
	state before the quench.} 
a quantity used extensively to study quantum quenches
and quantum chaos \cite{Peres, Jalabert, Goussev, Gorin}.  Though here we have shown it for initial pure states, however, this
identification is still valid for arbitrary initial mixed states as well. 
Furthermore, as we have shown previously in \cite{Pal:2023yik}, the fact that the CF of the WD is just the ACF
implies that this contains the information about the LC corresponding to the evolution  generated by the post-quench Hamiltonian,
and therefore, also determines the spread of the time-evolved wavefunction in the Hilbert space.
Here, our goal is to establish a similar relationship  between the CF and the ACF for the TPM scheme in the presence of the perturbation $W$.

\subsection{OTOC in a two-point measurement scheme as an  auto-correlation  function }\label{otocacf}
We now find out the CF for the TPM scheme described at the beginning of sec. \ref{TPM}. Substituting the distribution in Eq. \eqref{pdfo}
into the definition of the CF in Eq. \eqref{CF1} and using the resolution of the identity satisfied by the eigenstates of the operator 
$O$ we arrive at the relation \cite{Campisi}
\begin{align}\label{CF-OTOC}
& G(u, \tau)=\big<O_0\big| W_\tau^\dagger V^{\dagger}W_\tau V\big|O_0\big>~=F_{V,W}(\tau),\nonumber\\
& \text{where}~~V=e^{iu O}~.
\end{align}
This computation, therefore, shows that the CF of the distribution of $\Delta O$ in a TPM scheme is just the OTOC between the 
perturbation operator $W$ and $V=e^{iu O}$, where $O$ is the operator whose change is measured during the protocol. 
Since the OTOC is used to probe 
the scrambling of quantum information, the above identification indicates that CF of the distribution of $\Delta O$, 
and hence the probability distribution itself (which is related to the CF through a Fourier transform)  encodes the 
nature of information scrambling in a quantum system after it is subjected to a non-equilibrium protocol such as a quantum 
quench. Therefore, this helps to understand the scrambling of information from a thermodynamic perspective (this will be discussed
in the section \ref{OH0} below as well). For more details about the relation between the information scrambling, LE and statistics of work
done in chaotic quantum systems see \cite{Chenu1, Chenu2}.

We now show that by suitably re-writing the  relation in Eq.\eqref{CF-OTOC}, it can also be argued that the OTOC $F_{V,W}(\tau)$
also contains information about the LC of the evolution   generated by the operator $O$ itself, i.e., it determines the spread 
of a certain initial state in the Hilbert space corresponding to the operator $O$. To show this, we proceed as follows. 
First, from Eq. \eqref{CF-OTOC} it can be seen that we can write the CF as  
\begin{align}\label{CF-OTOC2}
	G(u, \tau) & =e^{iu O_0}\big<O_0,\tau\big| e^{-iu O} \big|O_0, \tau\big>~\nonumber\\
	& = e^{iu O_0}\big<O_0,\tau\big|O_0, \tau, u\big>~,
\end{align}
where we have defined $\big|O_0, \tau\big>=W_\tau \big|O_0\big>$. If $W$ is a unitary operator, this is just the state
at time $\tau$  evolved by the time-dependent operator $W_\tau$. Now we see that (apart from an overall phase factor) the
function $G(u, \tau)$ can be thought of as the (conjugate of) the ACF between the evolved state $\big|O_0, \tau, u\big>
=e^{-iu O} \big|O_0, \tau\big>$, and an initial state $\big|O_0, \tau\big>$. The state $\big|O_0, \tau, u\big>$ is therefore an evolved 
state in the second time of evolution $u$. Also as far as the evolution through  $u$ is concerned,  the initial state $\big|O_0, \tau\big>$
is time-independent. In this  picture,  we can think of the ACF corresponding to that of a sudden quench 
performed on an auxiliary system described by an auxiliary Hamiltonian $O$, such that the initial parameters of an 
operator $O_0$ are suddenly changed to a new set of values at $u=0$, and the subsequent evolution in $u$ is generated by the new 
auxiliary Hamiltonian $O$.

The discussion above indicates that the ACF, which is just the OTOC between $W_\tau$ and $V$, as well as the CF corresponding to the
distribution of the change of an observable $O$ in a TPM protocol, also has the information of the spreading of the initial state 
$\big|O_0, \tau\big>$ in the Krylov subspace generated by the operator $O$. As the next step, we  define the  SC
corresponding to the $u$ evolution, and  find out the associated Krylov basis and LCs.  This is what we describe in the next section.  

However, before moving on to the next section, here we note the following points. Firstly,  from now on, we shall neglect the overall
constant phase factor in front of the ACF in Eq. \eqref{CF-OTOC2}.  The effect of this phase factor is just 
to shift the average values of $\Delta O$, and does not have any extra physical meaning \cite{Pal:2023yik}.  In fact, this
phase factor can be set to unity by adjusting the lowest eigenvalue of $O$ to zero (i.e., $O_0=0$).

Secondly, as we elaborate upon below, 
the basic idea behind the notion of the SC is to write a `time'\footnote{Here, the time can be the real physical  time corresponding to the 
	Hamiltonian evolution, or it can be an auxiliary parameter conjugate to the eigenvalue of some observable (such as the parameter $u$ above). In the second case, we also call them  the circuit time, in analogy with the nomenclature used in the definition of the
	Nielsen and related related geometric measures of circuit complexity, where the evolution is generated in the circuit space by some suitable unitary operator.}
evolved state (generated by some hermitian operator, such as the Hamiltonian 
of a quantum system)  in terms of an orthogonal and complete basis, and find out the projection of the time-evolved state in terms 
of the elements of the basis. These projections, usually denoted as $\phi_n(u)$s, are just the probability amplitudes of obtaining the 
evolved state in each of these basis vectors. The SC is defined as the minimum (obtained in a special basis known as the Krylov basis) of the weighted sum of 
the modulus squared of $\phi_n$, and measures the 
spread of the evolved state in that basis. When the first state of the basis is the initial state at the start of the evolution ($u=0$), 
$\phi_0(u)$ is just the ACF. Therefore, for the evolution denoted in Eq. \eqref{CF-OTOC2}, we see that the OTOC is $\phi_0(u)$.
For the case where the Krylov basis has only two elements, the OTOC completely determines the spreading of the initial state.
However in the more general case with higher number of Krylov basis elements,  the OTOC between $W$ and $V$ does not have information 
about $\phi_n$s with $n\geq1$. In those cases, the SC of the $u$ evolution defined in the next section is more useful for studying the 
propagation of the initial state with circuit time, and, in a sense, has more information than standard OTOC.

Finally, we note that, in this paper we assume that the initial state of the system is a pure state (which we denote as $\big|O_0\big>$), so that
the CF associated with the distribution of Eq. \eqref{pdfo} can be directly  interpreted as an ACF. However, this conclusion can not be straightforwardly
extended to the cases where the initial state is a mixed state with density matrix $\rho_0$. To understand  this, we notice that 
for initial mixed states of the form $\rho_0=\sum_{n} p_{n}^0 \big|O_ n\big>\big<O_ n\big|$, the CF can be written as
\begin{equation}
	\begin{split}
   G(u, \tau)= \sum_n p_{n}^0 \big<O_n\big|  W_\tau^\dagger  e^{-i u O} W_\tau  e^{i u O} \big|O_ n\big>~\\ 
   =\sum_n p_{n}^0 e^{iuO_n} \big<O_n, \tau\big|e^{-i u O}\big|O_ n, \tau\big>~.
\end{split}
\end{equation}
Now it can be seen that $G(u, \tau)$ can not be directly written as an ACF (or Loschmidt amplitude). One way to proceed for such initial 
mixed states is to purify the  initial density matrix   by embedding it in a double-copy Hilbert space, so that, as in \cite{Chenu1}, the CF 
can be written as an ACF with respect to  the double-copy states. The resulting definition of SC in such cases is beyond the scope of the 
present paper, and we hope to return to this issue in a future work.

\section{Spread complexity associated with the  second time of evolution in a two-point measurement protocol}\label{SCTPM}

\subsection{Definition of the spread complexity of $u$-evolution}

Using the identification between the OTOC in the real time evolution and the ACF corresponding to the $u$ evolution
obtained in the previous section, here  we extend the definition of SC such that it captures the properties 
of the $u$-evolution of an initial state in a TPM scheme.

We first describe the Lanczos algorithm for constructing the Krylov basis and the subsequent definition of the 
SC of an initial state  $\big|O_0, \tau\big>$ under evolution generated by the operator $O$. 
The Krylov basis is used to write down the operator $O$ in a tri-diagonal form. 
In this construction, we start from the initial state $\big|\tilde{K_{0}}\rangle=\big|O_0, \tau\big>$,\footnote{Here, we have used an 
overall tilde in the notation for the Krylov basis elements to distinguish these from the Krylov basis generated through the Hamiltonian evolution.} i.e.,
we take the first state of the Krylov basis  is the initial state at $u=0$, and a new element 
of the  basis is obtained from the old ones as follows 
\begin{equation}\label{Krylov-basis}
	\big| \tilde{K}_{n+1} \big>=\frac{1}{\tilde{b}_{n+1}}\big[(O-\tilde{a}_{n})\big|\tilde{K}_n\big>-\tilde{b}_{n}\big|\tilde{K}_{n-1}\big>\big]~.
\end{equation}
The sets of coefficients $\tilde{a}_{n}$ and $\tilde{b}_{n}$ are the LCs\footnote{Once again we  have used an overall 
	tilde to distinguish these  from the LC generated in the Hamiltonian evolution.}, and these can be obtained from the moments of the 
ACF given in Eq. \eqref{CF-OTOC2} (see ref \cite{Viswanath}
for details of this procedure). The first set of coefficients $\tilde{a}_n$s  are given by  the 
expectation values of the operator $O$ in each of the Krylov basis elements
\begin{equation}\label{an}
	\tilde{a}_{n}=\langle {\tilde{K}_{n}|O|\tilde{K}_{n}}\rangle~,
\end{equation}
and the second set of coefficients $\tilde{b}_n$ are used to fix the normalisation of each $|\tilde{K}_{n}\rangle$ to unity.
We have to stop the recursion when $b_n=0$ at any particular step.
After obtaining the Krylov basis, we can expand the evolved state in terms of  this basis
\begin{equation}\label{expansion}
	\big|O_0, \tau, u\big>=\sum_{n}\tilde{\phi}_{n}(\tau,u)\big|\tilde{K_{n}}\rangle~,
\end{equation}
where the summation is over the dimension of the Krylov basis. Substituting this expansion in the Schrodinger-like equation satisfied by $\big|O_0, \tau, u\big>$, 
we obtain the following discrete equation satisfied by $\tilde{\phi}_{n}(u)$,
\begin{equation}\label{dse}
	i  \partial_u \tilde{\phi}_{n}(u)=\tilde{a}_{n}\tilde{\phi}_{n}(\tau,u)+\tilde{b}_{n}\tilde{\phi}(\tau,u)+
	\tilde{b}_{n+1}\tilde{\phi}_{n+1}(\tau,u)~.
\end{equation} 

To arrive at the associated notion of the SC, we first define the cost function $\mathcal{C}_{B}(u)=\sum_{n}n|\langle {O_0, \tau, u|B_{n}}\rangle|^2$, which can be thought of as a measure of the spreading of the time evolved state in an arbitrary
complete orthonormal basis $|B_{n}\rangle$. It was recently shown that \cite{Balasubramanian:2022tpr} 
for an evolution generated by a time-independent
hermitian operator, the basis which minimises this particular cost function is the Krylov basis constructed using that
hermitian operator. For our case, this operator is the observable $O$, and using the expansion in Eq. \eqref{expansion}
we arrive at the following definition of the SC in this case 
\begin{equation}\label{SCu}
	\mathcal{C}(\tau, u)=\sum_{n}n|\langle {O_0, \tau, u|\tilde{K}_{n}}\rangle|^2=\sum_{n}n |\tilde{\phi}_{n}(\tau,u)|^2~.
\end{equation}
The reason we have  also included the $\tau$ parameter  in the definition of the SC will be  explained below shortly.

\subsection{Implications of the  complexity of $u$-evolution}\label{significance}

With this definition of the SC of circuit time evolution, we now explore some of its consequences. First, notice that, though 
this definition of the SC is similar to the SC of usual Hamiltonian evolution, here
it also has information of the time evolution of the $W$ operator through the initial state $\big|O_0, \tau\big>$.
The more `complicated' the $W$ operator becomes with evolution through the system Hamiltonian $H$, the initial state of the circuit evolution  
gets more complicated.  Thus, both the system Hamiltonian for $t>0$, as well as the wing-flap 
perturbation operator  $W$,  influence the behaviour of the SC $\mathcal{C}(\tau,u)$ through the initial state $\big|O_0, \tau\big>$.
Therefore,  if we consider circuit evolution by the same operator $O$ of two initial states corresponding to the parameters,
say $\tau_1$ and $\tau_2$, depending upon their relative values the spreading of these two initial states can be 
very different.  This is the reason we have kept the parameter $\tau$ in the definition of SC in Eq. \eqref{SCu}, even though $\mathcal{C}(\tau, u)$
essentially measures the  complexity of spreading of an initial state with respect to the circuit time $u$.

In the following, we therefore consider the SC, $\mathcal{C}(\tau_i, u)$, for initial states with different values of the parameter $\tau_i$, which is the 
value of time where the perturbation $W$ has been applied on the system. Furthermore,  at this point, it is also 
useful to discuss the role played by the KC of the operator $W$, and the nature of the Hamiltonian $H$ on the initial state.
The KC is the analogue of the SC for the evolution of an operator in the Heisenberg picture and was 
introduced in \cite{Parker:2018yvk}. The procedure used to define it is very similar to the one used in Eq. \eqref{SCu}, i.e., one finds out 
the Krylov basis $|\tilde{K}_{n}\rangle$ generated
by the Liouvillian superoperator associated with the Hamiltonian generating the operator evolution in the Heisenberg picture (here
$W_\tau$ is such an operator), and expand the time-evolved operator (written as a state in the Hilbert space of the operator) in the Krylov basis
(analogous to Eq. \eqref{expansion}).  The weighted sum of the modulus squared of these expansion coefficients $\phi_{n}(t)$ defines the KC 
of the Heisenberg picture operator. 
Roughly, the more the operator spread in the Krylov basis under Hamiltonian evolution, the higher its Krylov complexity.

Here, in the $u$-evolution, the initial state on which the observable $O$ acts depends on one such time-evolved operator,
namely, the perturbation $W_\tau$. Therefore,  in some sense, it can be understood that the KC of the operator $W_\tau$ should 
affect the SC of the initial state $\big|O_0, \tau\big>$.

\subsection{Implications of the Lanczos coefficients and connection with thermodynamics }\label{OH0}
To understand the significance of the LCs corresponding to the circuit time evolution, in this section we assume that 
the observable under consideration $O$ is the Hamiltonian $H_0$ of the system. In this case, the initial state 
of the system before the $u$-evolution can be denoted as $\big|E_{00},\tau\big>$, where $E_{00}$ is the 
lowest eigenvalue of the Hamiltonian $H_0$, and  $G(u, \tau)$ represents the 
CF corresponding to the probability distribution of the 
change of the energy $\Delta E_0$ ($=E_{0m}-E_{00}$) of the initial system  due to the perturbation $W$.
Calculating the moments ($\tilde{M}_n$) of $G(u, \tau)$, we can obtain the averages of various powers of $\Delta E_0$, i.e., 
\begin{equation}
	\tilde{M}_n=\frac{d^n G(u, \tau)}{d u^n}\bigg|_{u=0}=(-i)^n \big<(\Delta E_0)^n\big>~.
\end{equation}
Since, $\tilde{M}_n$ are related to the LCs of the $u$-evolution, below we relate these averages with $\tilde{a}_n$ and $\tilde{b}_n$.

Using the expression for $G(u, \tau)$ given in Eq. \eqref{CF-OTOC}, we obtain the first two such averages to be equal 
to\footnote{As we have mentioned before, we have neglected an overall phase factor present in the CF during writing these formulas.}
\begin{equation}
	\begin{split}
	\big<\Delta E_0\big> = \big<E_{00}\big| W_\tau ^{\dagger} H_0 W_\tau \big|E_{00}\big>\\
	= \big<E_{00},\tau\big|H_0\big|E_{00},\tau\big>  ~,
   \end{split}
\end{equation}
and
\begin{equation}
	\begin{split}
			\big<(\Delta E_0)^2\big> &= \big<E_{00}\big|[ W_\tau, H_0]^{\dagger}[ W_\tau, H_0]\big|E_{00}\big>\\
		&= \big<E_{00},\tau\big| H_0^2\big|E_{00},\tau\big>~.
	\end{split}
\end{equation}
Now from the identification between the moments of the CF and the LCs in a quench scenario made in \cite{Pal:2023yik}, we obtain the 
first two LCs in terms of the above averages as 
\begin{equation}
	\tilde{a}_0=\big<(\Delta E_0)\big>=\big<E_{00},\tau\big|H_0\big|E_{00},\tau\big>~,
\end{equation}
\begin{equation}
	\begin{split}
	\tilde{b}_1^2 &=\big<(\Delta E_0)^2\big>-(\big<\Delta E_0\big>)^2\\
	&= \big<E_{00},\tau\big| H_0^2\big|E_{00},\tau\big> - \big<E_{00},\tau\big|H_0\big|E_{00},\tau\big>^2~.
\end{split}
\end{equation}
Therefore, we see that $\tilde{a}_0$ is the average of the initial Hamiltonian $H_0$ in the initial state, while 
$ \tilde{b}_1^2$ is the  variance of $ \Delta E_0 $. Similar relations 
can also be established between the higher order LC and various powers of averages of $\Delta E_0$. Analytical 
forms for these are complicated, and therefore, we do not present them here for brevity. 

Here we also note that, $ \tilde{b}_1$, in the form written above, is very similar to the Fubini-Study (FS) metric
	generated by $H_0$ from the initial state $\big|E_{00},\tau\big>$ of the circuit time evolution.\footnote{More specifically,  for each 
	value of the circuit time, $ \tilde{b}_1$ is equal to the FS metric. }
	This line element is the starting point of the definition of an alternative geometric notion of circuit complexity compared to that of Nielsen's
	\cite{Nielsen1, Nielsen2, Nielsen3, Jefferson:2017sdb},
	and is known as the FS complexity \cite{Chapman:2017rqy}.   In this definition of the circuit complexity, one uses the FS
	line element defined on the space of pure states as a state-dependent measure of the cost function, and subsequently obtains the associated 
	complexity between two states by finding out the geodesic distance between them as measured by the FS 
	metric \cite{Bueno:2019ajd} (see also \cite{tapo1,tapo2}).\footnote{For works that use the related so-called quantum information metric to define the FS complexity in the parameter space 
		of  quantum many-body systems 
		showing ground and exited state quantum phase transitions see, \cite{FS1, FS2, LMG}. }
	In a similar vein, we see that 
	another state-dependent cost $F_{|\langle H_0 \rangle |}= |\big<E_{00},\tau\big|H_0\big|E_{00},\tau\big>|$ proposed in \cite{Magan}, is equal to the
	modulus of the LC $\tilde{a}_0$, with the role of the instantaneous Hamiltonian played by $H_0$.

The fact that $\tilde{a}_0$ and $ \tilde{b}_1$ are related to the well known cost functions in geometric approaches to the circuit complexity
	provides a possible way of connecting these measures of 
	circuit complexity  and the SC studied here. In principle, by using the reverse argument, one can define other new types 
	of  cost functions from other LCs as well, and study the resulting measures of  circuit complexity. Therefore, we can conclude that 
	it is possible to understand  various cost functions as LCs with respect to some unitary evolution.
	Hopefully, this will shed new light on the connection between these two distinct measures of defining  complexity of quantum systems.

\subsection{Connection with the Fidelity OTOC}

Before moving on to the following sections, where we compute the SC of the $u$-evolution and the associated quantities, here we discuss an interesting connection with the Fidelity OTOC (FOTOC), a class of 
OTOCs, where one sets the operator $V$ in Eq. \eqref{OTOC}, as the projector on the initial state \cite{Garttner1}.
This quantity can provide important insights into the scrambling of information in quantum many-body systems.
Recently, in \cite{fotoc},  this quantity has been studied for the Dicke model of quantum optics and 
it was shown there that FOTOCs can connect scrambling, volume-law of the Renyi entropy, and thermalisation.
Specifically, here  we show that for small values of the evolution parameter, the LC  $\tilde{b}_{1}$ is related  to the FOTOC.

The FOTOC is defined as  (according to the notation used in this paper)
\begin{equation}\label{foto}\
	\mathcal{F}(t) = \big<O_0\big| W_t^\dagger \rho (0) W_t \rho(0) \big|O_0\big>~,
\end{equation}
where $\rho(0) =\big|O_0\big> \big<O_0\big|$ is the projector on the initial state, and $W_t$ is the Heisenberg 
picture operator corresponding to $W$. To understand the significance of this quantity, we consider the 
modulus squared of the CF of the TPM protocol (given in Eq. \eqref{CF-OTOC}),
\begin{equation}\label{proba1}
	|G(u, \tau)|^2=\big|\big<O_0\big| W_\tau^\dagger e^{-iu O}W_\tau e^{iu O}\big|O_0\big>\big|^2~.
\end{equation}
With a little manipulation we can rewrite this  as 
\begin{equation}\label{fotoc2}\
	|G(u, \tau)|^2 = \big<O_0\big| \tilde{W}_t^\dagger \rho(0) \tilde{W}_t \rho(0) \big|O_0\big>~=\mathcal{F}(t,u)~,
\end{equation}
where, $\tilde{W}_\tau = W^{\dagger}_{\tau} e^{i u O }  W_{\tau}  $. As the last identification indicates, 
this is just the FOTOC between  the operators $\tilde{W}_\tau$ and $V=\rho(0)$.
Notice that $|G(u, \tau)|^2$ is also the modulus squared of the Fourier transform of the probability
distribution of $\Delta O$ is a TPM protocol (see Eq. \eqref{pdfo}), and it can also be written as the product of Fourier transforms
of two probability distributions of two different  TPMs, one for the measurement of the operator $O$ (corresponding to the 
change of eigenvalue $\Delta O=O_n-O_0$), and the other for the measurement of the operator $\tilde{O}=-O$ (and hence, 
corresponding change of eigenvalue is $\Delta \tilde{O}=-\Delta O=O_0-O_n$). Furthermore, from the expression for $|G(u, \tau)|^2$
in Eq. \eqref{proba1}, we see that, when the OTOC of the TPM is interpreted as the ACF, the above FOTOC is also the survival 
probability of the initial state $\big|O_0, \tau\big>$ under $O$ evolution.

Using the survival probability in Eq. \eqref{proba1}, we can find out the behaviour of the FOTOC at large $u$ by computing its long-$u$ 
average 
\begin{equation}
	\bar{\mathcal{F}}(\tau):= \lim_{\mathcal{U} \rightarrow \infty} \frac{1}{\mathcal{U}}
	\int_{0}^{\mathcal{U}} \text{d}u \mathcal{F}(\tau,u)~.
\end{equation}
Using the expression for $G(u, \tau)$ from Eq. \eqref{CF1} as well as Eqs. \eqref{distribution} and \eqref{pdfo}, we obtain  this to be
\begin{equation}
\bar{\mathcal{F}}(\tau)= \sum_{n}\big|\big<O_n\big|W_\tau\big|O_0\big>\big|^4~.
\end{equation}
Therefore, here the long-$u$ average of the FOTOC is just the sum of the square of the transition probabilities. Furthermore, noting that 
$\big|O_0, \tau\big>=W_\tau \big|O_0\big>$, we also see that $\bar{\mathcal{F}}(\tau)$ is just the IPR of the state $\big|O_0, \tau\big>$ 
in the eigenbasis of the operator $O$ (see section \ref{iprb1} below).

We can obtain an instructive  behaviour of the FOTOC for small values of  $u$ as well.
For a TPM protocol where $O=H_0$, the FOTOC becomes the survival probability of the state 
$\big|E_{00},\tau\big>$ under the Hamiltonian evolution
\begin{equation}\label{foto2}\
	\mathcal{F}(\tau,u) = \big| \big<E_{00},\tau \big| e^{-i H_{0} u}  \big|E_{00},\tau\big>\big|^2,
\end{equation}
so that, for $u<<1$, we can write  it as 
\begin{equation}
	\begin{split}
			\mathcal{F}(\tau,u) &= 1 - u^{2} \Big(\big<E_{00},\tau\big| H_0^2\big|E_{00},\tau\big> - \big<E_{00},\tau\big|H_0\big|E_{00},\tau\big>^2 \Big)  \\
		&= 1 - u^2 \tilde{b}_{1}^{2}~.
	\end{split}
\end{equation}
Therefore, for small values of $u$, the FOTOC decays quadratically with $u$, and the decay rate is characterised by the LC $\tilde{b}_{1}$.
For this kind of quadratically decaying survival probability, it is known that the corresponding SC would grow quadratically 
with $u$ \cite{SCint}.

We also notice that, $\tilde{b}_{1}$ is equal to the quantum Fisher information of $H_0$ in the time-evolved state. 
Quantum Fisher information for a pure state is defined as the variance of $H_0$ \cite{Toth}, and is a central quantity 
in parameter estimation as well as a key witness for multipartite entanglement \cite{Pezze}.


\section{Spread  complexity of $u$-evolution associated with  Lie algebras}\label{algebra}

In this section we analytically obtain the SC of the $u$-evolution by using some assumptions about the operator $O$ and the Hamiltonians
generating the time evolution.  Specifically, we assume that $O$ is an element of some Lie algebra, and use the analytical technique developed in
\cite{Caputa1,Balasubramanian:2022tpr} to obtain the Krylov basis and the complexity in Eq. \eqref{SCu}.
To complement these analytical computations, in the next section,  we numerically obtain the SC for realistic spin systems in both integrable and chaotic cases.
These examples will help us to clearly understand various properties of the quantity $\mathcal{C}(\tau, u)$ we have introduced above.

\subsection{The $u$-evolved state }
Assuming that $W$ is an unitary operator, we first write down the ACF in Eq. \eqref{CF-OTOC2} in the following way,
\begin{equation}\label{auto}
	\begin{split}
	G(u, \tau)&=\big<O_0\big| e^{-i W_{\text{ef}}(\tau)}  e^{-iu O} e^{i W_{\text{ef}}(\tau)} \big|O_0\big>\\
	&=\big<O_0\big|  e^{-iu O_{\text{ef}}(\tau)}  \big|O_0\big>~,
\end{split}
\end{equation}
where we have used the notation
\begin{equation}\label{effeWO}
	W_{\tau}= e^{-i W_{\text{ef}}(\tau)}~, ~~~\text{and}~~~ O_{\text{ef}}(\tau)=e^{-i W_{\text{ef}}(\tau)} O e^{i W_{\text{ef}}(\tau)}~.
\end{equation}
As is evident from these relations, $W_{\text{ef}}(\tau)$ is a hermitian operator which contains the information of the time evolution
of the operator $W$ under the Hamiltonian $H$, and similarly, $O_{\text{ef}}$ is also a hermitian operator (since $O$ is also hermitian)
which encodes  the effect of $W_{\text{ef}}(\tau)$ on the observable $O$. 

From Eq. \eqref{auto} we see that, $G(u, \tau)$ can be equivalently viewed as the ACF between the $u$-evolved state $e^{-iu O_{\text{ef}}}  \big|O_0\big>$
and the initial state $  \big|O_0\big>$. This is an alternative viewpoint from the one described in the previous section, with the  most important 
difference between the two scenarios being that, in the present case, the initial state of the circuit time evolution is  actually independent of
the time $\tau$ where the perturbation is applied. In fact here, the initial state is just an eigenstate of the operator $O$. Even so, the 
$u$-evolved state is still non-trivial, since the $u$-evolution here is generated through the  `effective' observable $O_{\text{ef}}(\tau)$
rather than $O$ itself (unlike the previous scenario), and the operator $O_{\text{ef}}(\tau)$ now is time-dependent (in the previous case the initial 
state was time-dependent). This is the viewpoint we use throughout the  present section and will return to the previous version 
in the next section, though we emphasize that both of them are equivalent (since they are just alternative ways of writing 
the same quantity --  an OTOC), and which one has to be used is just a matter of convenience (we discussed this equivalence  briefly in 
Appendix \ref{interpretations}). 

To proceed analytically, and to use the geometric method of obtaining the LCs and the Krylov basis developed in \cite{Caputa1, Balasubramanian:2022tpr}, 
we assume that the operators $W_{\text{ef}}(\tau)$ and $O$  are of the following form
\begin{equation}
W_{\text{ef}}(\tau)= \alpha f(\tau) \big(K_++K_-\big)~=f(\tau)L, ~~~~\text{and} ~~~~  O=\gamma K_0~,
\end{equation}
where $\alpha$ and $\gamma$ are two real constants, $f(\tau)$ is a real function of time, and 
the three operators $K_{i}$ are assumed to be the generators of a Lie algebra. In this paper, we consider the cases when the Lie algebra under consideration 
is either $su(1,1)$ or $su(2)$, so that the generators satisfy the following commutation relations
\begin{equation}
	\big[K_-,K_+\big]= 2 \sigma K_0~,~
	~\big[K_0,K_{\pm}\big]= \pm K_{\pm}~.
\end{equation}
When $\sigma$ is $1$, the algebra generated by these operators is a $su(1,1)$ algebra, while, for $\sigma=-1$ they become the generators
of the $su(2)$ algebra. Note that we have assumed that the operator $O$ is only proportional to $K_0$. It is, of course, possible to take 
a more general form for both the operators $W_{\text{ef}}$ and $O$ (e.g., a general combination 
of all three generators),\footnote{In that case, the Krylov basis would  not simply be the basis states of the representation of the 
Lie algebra (or at most related through some phase factor, as is the case below), rather would be given by linear combinations of them.}
however, for our purposes, this relatively simple form is sufficient. 

To evaluate the final operator $O_{\text{ef}}(\tau)$, we need the expressions for the commutator $[L, O]$ as well as  the higher order nested 
commutators between $L$ and  $[L, O]$.  From the definitions of $L$ and $O$, we first obtain 
\begin{equation}
	[L, O] = \alpha \gamma (K_--K_+)=\gamma \bar{L}~,~~~  [L,[L, O] ]=-4 \alpha ^2 \sigma O~,
\end{equation}
where for convenience, we have renamed $\alpha (K_--K_+)$ as $\bar{L}$.  Similarly, all the higher-order nested commutators can be evaluated 
in terms of $\bar{L}$ and $O$. Now using the Baker–Campbell–Hausdorff lemma,  we can evaluate $O_{\text{ef}}(\tau)$ from Eq. \eqref{effeWO} to be the following series
\begin{equation}
	\begin{split}
	O_{\text{ef}}(\tau) = O \Big[1-\frac{f(\tau)^2}{2 !} (-4 \alpha ^2 \sigma)+\frac{f(\tau)^4}{4 !} (-4 \alpha ^2 \sigma)^2+\cdots\Big]-\\
	i \gamma \bar{L}\Big[f(\tau) -\frac{f(\tau)^3}{3 !} (-4 \alpha ^2 \sigma) +\frac{f(\tau)^5}{5 !} (-4 \alpha ^2 \sigma)^2+\cdots\Big]~.
\end{split}
\end{equation}
This series has different expressions for $su(2)$ and $su(1,1)$ algebras. For the $su(2)$ algebra (for which $\sigma=-1$), taking $\gamma=2\alpha$
we can sum the above series and write the final expression in a compact form as 
\begin{equation}\label{Oe2}
	O_{\text{ef}}(\tau) =  O \cos (2 \alpha f(\tau)) - i \bar{L} \sin (2 \alpha f(\tau))~.
\end{equation}
On the other hand, for the $su(1,1)$ algebra (for which $\sigma=1$), once again taking  $\gamma=2\alpha$, we arrive at the following  expression for the `effective observable' operator
\begin{equation}\label{Oe11}
	O_{\text{ef}}(\tau) =  O \cosh (2 \alpha f(\tau)) - i \bar{L} \sinh (2 \alpha f(\tau))~.
\end{equation}
From these expressions for the effective operator $O_{\text{ef}}(\tau)$, we see that for both the Lie algebras under consideration, this 
is a general element of the respective algebra, and, therefore,  can be written in the following general form
\begin{equation}
	O_{\text{ef}}(\tau) = \mathcal{A}_0(\tau) K_0 + i \mathcal{A}_1(\tau)  (K_+-K_-)~, 
\end{equation}
where the $\tau$ dependent coefficients $\mathcal{A}_0(\tau)$ and $\mathcal{A}_1(\tau)$ can be read off from Eqs. \eqref{Oe2} and \eqref{Oe11}. For  the $su(2)$ algebras, these are given  by 
\begin{equation}
	\mathcal{A}_0=2\alpha \cos (2 \alpha f(\tau))~,~~~\mathcal{A}_1= \alpha \sin (2 \alpha f(\tau)) ~,~~
\end{equation}
while for the  $su(1,1)$ algebra we have 
\begin{equation}
	\mathcal{A}_0=2\alpha \cosh (2 \alpha f(\tau))~,~~~\mathcal{A}_1= \alpha \sinh (2 \alpha f(\tau)) ~.
\end{equation}

\subsection{The Lanczos coefficients and the Krylov basis}

For the $u$-evolution of Eq. \eqref{auto}, the operator $O_{\text{ef}}$
plays the role of the Hamiltonian, and the corresponding Krylov basis vectors are generated by the action of this operator on the initial state  $\big|O_0\big>$.
As we have discussed at the beginning of this section, though the interpretation of the ACF of \eqref{CF-OTOC} used in the present section 
is slightly different from the previous section, for convenience, we still  denote the Krylov basis and the 
LC with tildes to distinguish them from 
those corresponding to the Hamiltonian evolution. In particular, with the notation we use here,
the action of the operator $O_{\text{ef}}$ on the Krylov basis is of  the following form
\begin{equation}\label{OKrylov-basis}
	O_{\text{ef}} \big|\tilde{K_n}\big> = \tilde{a}_{n}\big|\tilde{K_n}\big>+\tilde{b}_{n+1}\big| \tilde{K}_{n+1} \big>
	+\tilde{b}_{n}\big|\tilde{K}_{n-1}\big>~.
\end{equation}

Now since $O_{\text{ef}}(\tau)$ is an element of a Lie algebra, the  $u$-evolved state $ \big|\Psi(u)\big>=e^{-iu O_{\text{ef}}}  \big|O_0\big>$
is a generalised or Perelomov coherent state (CS) associated with  $SU(2)$ (or $SU(1,1)$) Lie group \cite{Perelomov}.  Therefore, we can use the geometrical
method based on the generalised CS developed in \cite{Caputa1, Balasubramanian:2022tpr} to directly obtain the Krylov basis and the associated LCs.
We obtain these  separately for the two algebras under consideration.

\textbf{ Case-1: the $su(2)$ algebra.}
First, we specify the action of the  generators of  $su(2)$ algebra on the basis for representation $\big|j, -j+n\big>$.  These are given by the standard
formulas\footnote{To be consistent with the notations used in \cite{Caputa1}, we have shifted $n \rightarrow j+n$ from the usual convention, 
	e.g., the one used in \cite{Ban} to derive the decomposition formulas associated with the $su(2)$ Lie algebra.}
\begin{equation}\label{su21}
\begin{split}
	K_0 \big|j, -j+n\big> &= (-j+n)~ \big|j, -j+n\big> \\ 
	K_+ \big|j, -j+n\big> &= \sqrt{(n+1)(2j-n)}~\big|j, -j+n+1\big> \\ 
	K_- \big|j, -j+n\big> &= \sqrt{n(2j-n+1)}~\big|j, -j+n-1\big>~,
\end{split}
\end{equation}
where $j=0, 1/2, 1, \cdots$ and $n=0, 1, \cdots, 2j$. Furthermore, the conditions $K_+ \big|j, j\big> =0$, and $K_- \big|j, -j\big> =0$
are  satisfied. Here we assume that the initial state $\big|O_0\big>$ is the state $\big|j, -j\big>$, and therefore, is annihilated by the operator $K_-$.

Now comparing the action of the operator $O_{\text{ef}} $ on the   states $\big|j, -j+n\big>$  of  the above representation 
(using the relations in Eqs. \eqref{su21}), and comparing with 
the definition of the Krylov basis in  Eq. \eqref{OKrylov-basis}, we get the elements of the Krylov basis and the LCs in this case 
to be 
\begin{equation}
	\begin{split}
	\big|\tilde{K_n}\big>&=i^{n+1}\big|j, -j+n\big>~, ~~~ \tilde{a}_{n}=\mathcal{A}_0 (n-j)~, \\
	\tilde{b}_{n} &=\mathcal{A}_1 \sqrt{n(2j-n+1)}~,
\end{split}
\end{equation}
where, from Eq. \eqref{Oe2} we get the coefficients $\mathcal{A}_0= 2 \alpha \cos (2 \alpha f(\tau))$ and $A_1=\alpha  \sin (2 \alpha f(\tau))$. 
Notice the extra phase factor of  $i^{n+1}$ in front of the element of the Krylov basis above, which usually remains absent from the Krylov basis
generated by the Hamiltonian evolution  \cite{Balasubramanian:2022tpr}. This is due to the  minus sign between the operators $K_+$ and $K_-$ 
in the  expression for $O_{\text{ef}} $, and does not have any effect in the SC.

\textbf{ Case-2: $su(1,1)$ algebra.} In this case the action of the generators on the basis of representation $|h,n \rangle$ are given by the 
following relations
\begin{equation}\label{su111}
\begin{split}
	K_0 |h,n \rangle &= (h+n)~ |h,n \rangle\\
	K_+ |h,n \rangle &= \sqrt{(n+1)(2h+n)}~\big|h, n+1\big>\\
	K_- |h,n \rangle &= \sqrt{n(2h+n-1)}~\big|h, n-1\big>~,
\end{split}
\end{equation}
here $n$ is a non-negative integer, and the constant $h$ is called the Bargmann index. It is well known that, for a single-mode bosonic 
representation of $su(1,1)$ algebra, the Bargmann index can take values $1/4$ or $3/4$ (see, e.g.,  \cite{Gerry:91}). In this paper,
we assume that the basis corresponding to a unitary irreducible representation of the $su(1,1)$ 
Lie algebra is a set of states which contains   an even number of bosons, so that $h$ is taken to be  $1/4$. Here we also assume that the initial state 
is $\big|O_0\big>=\big|h,0\big>$.

Once again, considering the action of the operator $O_{\text{ef}}(\tau)$ on the basis $|h,n \rangle$, using Eqs. \eqref{su111}, 
and comparing the result with the definition in Eq. \eqref{OKrylov-basis}, we obtain the  Krylov basis and LCs to be 
\begin{equation}
	\begin{split}
	\big|\tilde{K_n}\big>&=i^{n+1}|h,n \rangle~, ~~~ \tilde{a}_{n}=\mathcal{A}_0 (n+h)~, \\
	\tilde{b}_{n}&=\mathcal{A}_1 \sqrt{n(2h+n-1)}~,
\end{split}
\end{equation}
where from Eq. \eqref{Oe11} we now have $\mathcal{A}_0=2 \alpha \cosh (2 \alpha f(\tau))$, and $\mathcal{A}_1=\alpha  \sinh (2 \alpha f(\tau))$.

Since the LCs are dependent on $\tau$ through the coefficients $\mathcal{A}_0$ and $\mathcal{A}_1$, it is 
instructive to  compare the magnitudes of LC for different fixed values of $\tau=\tau_i$.\footnote{We analyse a similar setting
numerically in the next section, where the different values $\tau$ fix the initial state of the $u$-evolution. Here, the values of $\tau$ 
changes the coefficients of  the generators in the operator $O_{\text{ef}} (\tau)$ generating the $u$-evolution with respect to the
initial state $\big|O_0\big>$. } For the $su(2)$ algebra, the LC changes periodically with $\bar{\tau}=f(\tau)$, while for the 
$su(1,1)$ algebra they grow with $\bar{\tau}$, with the growth being exponential for large values of $\bar{\tau}$.
Furthermore, for large $\bar{\tau}$, the growth rates of both sets of coefficients $\tilde{a}_{n}$ and $\tilde{b}_{n}$ with respect to $\bar{\tau}$ 
are equal and are fixed by the constant $\alpha$.

\subsection{Evolution of spread complexity}
Using the LCs obtained above, we now find out the SC of the $u$-evolution defined in Eq. \eqref{SCu}. Here, we shall show the computation of SC only for the 
$su(1,1)$ algebra, and an entirely similar procedure can be followed to find out the SC for the $su(2)$ algebra.

First we use the decomposition formula for the  $su(1,1)$ algebra (see e.g., \cite{Ban}) to write the $u$-evolved state in the following form
(using $h=1/4$)
\begin{equation}
	\begin{split}
	 \big|\Psi(u)\big>&=\exp (C_{+}K_+) \exp (C_{0}K_0) \exp (C_{-}K_-)\big|h,0\big>\\
	 &=(C_0)^{1/4}\sum_{n=0}^{\infty} \frac{1}{n !}  (C_{+})^n (K_+)^n \big|h,0\big>~,
 \end{split}
\end{equation}
where,  $C_{\pm}$, $C_0$ are functions of $\tau$ and $u$, and are given by the expressions 
\begin{equation}
	C_{\pm} (\tau,u)= \frac{c_{\pm}(\tau,u)}{g(\tau,u) \Theta (u)} \sinh \Theta(u)~,~~C_{0}(\tau,u)=g(\tau,u)^{-2}~,
\end{equation}
with
\begin{equation}
\begin{split}
		g(\tau)&=\cosh \Theta (u)-\frac{c_{0}(\tau,u)}{2 \Theta (u)} \sinh \Theta(u)~,~\\
		c_{+}&=- c_{-} = u  \alpha \sinh (2 \alpha f(\tau))~,~c_{0} = -2iu \alpha \cosh (2 \alpha f(\tau))~\\
		\Theta (u)&=\big[\Big(\frac{c_{0}}{2}\Big)^2-c_+c_-\big]^{1/2}~=iu \alpha~.
	\end{split}
\end{equation}
From the above expression for the evolved state we get the simplified expression for the ACF to be $\mathcal{S}(u)=\langle \Psi(u)\big| h,0 \rangle=(C_+ (\tau))^{1/4}$, and comparing the expansion in second line  above with the expansion in Eq. \eqref{expansion} of an arbitrary $u$-evolved state in the Krylov basis we obtain
\begin{equation}
	\begin{split}
	\tilde{\phi}_{n}(\tau,u)=\mathcal{N}_n  (C_0 (\tau,u))^{1/4} (C_{+}(\tau,u))^n~, ~\\~~\text{where}~~
	\mathcal{N}_n=i^{-(n+1)}\sqrt{\frac{\Gamma(n+\frac{1}{2})}{n!\sqrt{\pi}}}~.
\end{split}
\end{equation}
Using these expressions for $\tilde{\phi}_{n}(\tau,u)$, we can perform the summation in Eq. \eqref{SCu} exactly, and the resulting expression for the SC of $u$-evolution can be compactly written as \cite{Balasubramanian:2022tpr, Pal:2023yik}
\begin{equation}
\begin{split}
& \mathcal{C}(\tau, u)= \frac{|\tilde{\phi}_{1}(\tau,u)|^2}{\big(1-F(\tau, u)\big)^{3/2}}~,\\ &\text{where}~~~
   F(\tau, u)=|C_{+} (\tau,u)|^2~.
\end{split}
\end{equation}

\section{Spread complexity of circuit evolution in integrable and chaotic systems}\label{spinchain}

In this section we numerically study the SC of circuit evolution for different integrable and chaotic 
interacting quantum systems using  the following return amplitude
\begin{equation}\label{ampl0}
\begin{split}
	G(u,\tau) &= \big< E_{0}\big| W_{\tau}^{\dagger} e^{- i H_{0} u}  W_{\tau} \big|E_{0}\big>~\\
	&=\big< E_{0}, \tau\big|  e^{- i H_{0} u} \big|E_{0}, \tau\big>~,
\end{split}
\end{equation}
where $W$ is some local operator acting on $\big|E_{0}\big>$, a bulk eigenstate of $H_0$, and $ W_{\tau} = e^{- i H_1 \tau} W e^{ i H_1 \tau} $
is the Heisenberg operator at time $\tau$ evolved through the post-quench Hamiltonian $H_1$. Therefore, the observable $O$ that one 
measures in the TPM protocol is the Hamiltonian $H_0$ of the system before $t<0$. This is the case we considered in section \ref{OH0}
to understand the significance of the LC.  Notice that, since here $H_0$ is a Hamiltonian, the parameter $u$ can be thought of as the Fourier 
conjugate of the eigenvalues of $H_0$. In fact, in the absence of the perturbation $W$, the resulting ACF would just represent the time evolution
of an eigenstate of $H_0$. The presence of the perturbation changes the initial state $\big|E_{0}\big>$ non-trivially, so that finding the spreading 
under evolution generated by $H_0$ has a well defined meaning, as it carries the  implications of the perturbation $W$ in the 
evolution of the TPM protocol. 

\begin{figure}
	\centering
	\centering
	\includegraphics[width=0.5\textwidth]{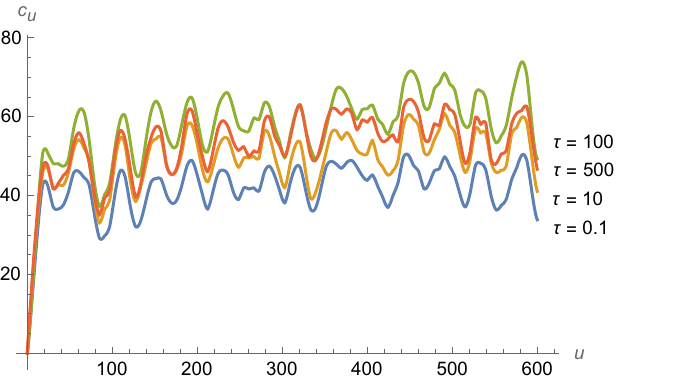}
	\caption{SC of the $u$-evolution when both $H_0$ and $H_1$ are integrable, for different fixed values of the parameter $\tau$, which fixes the initial state. Here, $H_{0}=H_{\text{int}}(1,0.4)$ and $H_{1}=H_{\text{int}}(1,0.7)$ respectively.}
	\label{fig:SC_int_int}
\end{figure}

\begin{figure}
	\centering
	\includegraphics[width=0.5\textwidth]{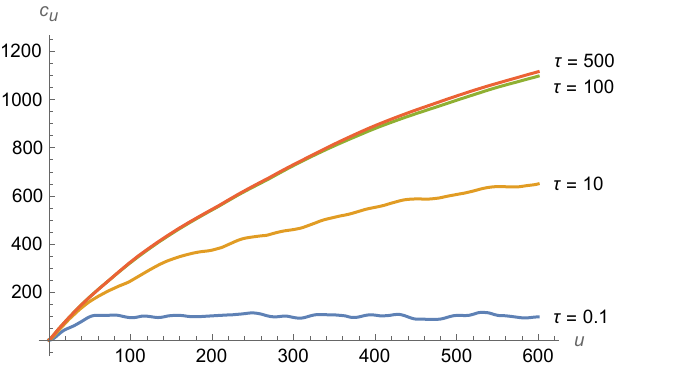}
	\caption{SC of the circuit evolution for different values of $\tau$,  when $H_0$ is integrable and $H_1$ is chaotic. Here, $H_{0}=H_{\text{int}}(1,0.4)$ and $H_{1}=H_{\text{cha}}(1,1.4,-0.6)$.}
	\label{fig:int_chao}
\end{figure}

\begin{figure}
	\centering
	\includegraphics[width=0.5\textwidth]{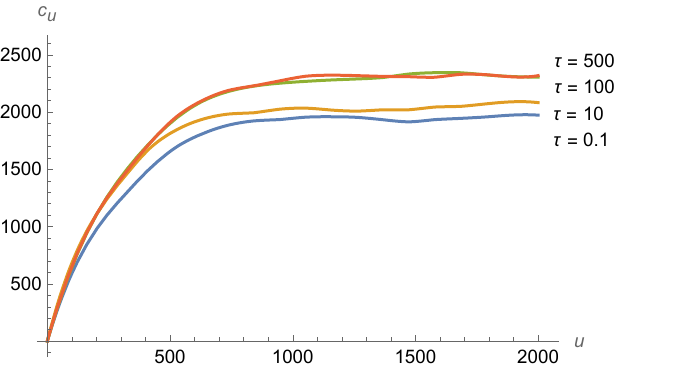}
	\caption{SC of the circuit evolution for different values of $\tau$,  when $H_0$ and $H_1$ both chaotic. Here, $H_{0}=H_{\text{cha}}(0.8,1.2,-0.6)$ and $H_{1}=H_{\text{cha}}(1,1.4,-0.6)$.}
	\label{fig:chao_chao}
\end{figure}

Here we  take $H_0$ and $H_1$ to be  the integrable or chaotic limit of the Ising chain 
with different combinations, as we describe below.  In all the cases 
considered below, we take the perturbation $W$ to be $W=e^{-i \theta \sigma_z}$ with $\theta=\pi/2$.
The explicit form for the Hamiltonians we consider are the following :
\begin{equation}\label{Hint}
	H_{\text{int}}(J,h) = -\sum _{i=1}^N \Big [J \sigma _{i+1}^z \sigma_i^z +  h \sigma _i^x \Big]~,
\end{equation}
and 
\begin{equation}\label{Hchao}
	H_{\text{cha}}(J,h,g) = -\sum _{i=1}^N  \Big[J \sigma _{i+1}^z \sigma_i^z  + h \sigma _i^x +  g \sigma _i^z  \Big]~,
\end{equation}
where the $\sigma_i$ are the Pauli matrices at the $i$-th site of the chain. In the first case, $g=0$, and the Hamiltonian is that
of an Ising model with a transverse field and is integrable \cite{Izrailev}. On the other hand, in the second case, with non-zero values of 
both the parameters $g$ and $h$, the Hamiltonian is non-integrable, as can be verified by finding out the level spacing distribution
of $H_{\text{cha}}(J,h,g)$, which in this case is close to the  Wigner-Dyson distribution  characterising  a  chaotic system \cite{Banuls}.
In all the cases considered below, we take $N=12$. The chosen parameter values are indicated in the captions of figures
\ref{fig:SC_int_int} -- \ref{fig:chao_chao}.

\textbf{Case-1.}
First we consider the case when both  $H_0$ and $H_1$ are taken as  integrable Hamiltonians. In this case the SC, $\mathcal{C}(\tau_i,u)$ 
(defined in Eq. \eqref{SCu} above) for different fixed values of $\tau=\tau_i$ are plotted in Fig. \ref{fig:SC_int_int}. As we have explained in section 
\ref{significance} above, these values  of $\tau$ fix the initial state $\big|O_0, \tau\big>$ of the $u$-evolution. In the TPM protocol, the parameter 
$\tau$ denotes the time when the perturbation $W$ is applied, and therefore, starting from the lower values 
of $\tau$, its increasing values 
indicate  that the initial state   $\big|O_0, \tau\big>$ gets more and more `complicated' with $\tau$. Furthermore, how this state 
changes with time is  encoded in the time-evolved operator $W_\tau$, and a useful measure of this is the operator  complexity $\mathcal{C}_{W}$ 
of the  perturbation operator $W$.

From fig. \ref{fig:SC_int_int} we observe that $\mathcal{C}(\tau_i,u)$ does not have a prominent growth at early times and does not show saturation, 
even for large values of $u$.  Similarly, from the plots of $\mathcal{C}(\tau_i,u)$ for different values of $\tau_i$, we see that, as one varies $\tau$, even 
for large $\tau$, the SC does not attain steady behavior. These features of SC evolution can be attributed to the fact that $H_0$, generating the circuit time
evolution  is integrable, so that the SC  shows the usual oscillatory behaviour observed for an integrable Hamiltonian.

\textbf{Case-2.}
Next, we consider the case when $H_0$ is the integrable Hamiltonian of Eq. \eqref{Hint}, and $H_1$ is the chaotic one given in Eq. \eqref{Hchao}. 
In this case  we plot  $\mathcal{C}(\tau_i,u)$ for different values of $\tau$ in Fig. \ref{fig:int_chao}. Comparing with Fig. \ref{fig:SC_int_int} 
(where $H_1$ was taken as integrable) we see that, in this case the initial growth of the  SC is more  prominent, and the differences 
in the magnitudes of  $\mathcal{C}(\tau_i,u)$ for different values $\tau_i$ are greater (until at later times, when the operator 
complexity of $W$ saturates and initial states for different large values of $\tau$ are similar). This can be understood from the 
fact that here the post-quench Hamiltonian $H_1$, which determines the initial state of the $u$-evolution, is in fact a chaotic one, so that
even for small values of $\tau$, the initial states are quite different from each other. On the other hand, at later times $\tau_i$, the  profile of
$ \mathcal{C}(\tau_i,u)$ are quite similar, since $H_1$ being chaotic, the $W$ evolution gets saturated at these values of $\tau_i$. Thus, the pattern of SC
for different  $\tau$ occurs due to the differences in growth of the perturbation operator $W$ under chaotic and integrable dynamics in the present
case and in case-1, discussed in the previous paragraph. 
This is reflected in the dependence of $\tilde{b}_1$ on OTOC and Fidelity OTOC (discussed at the end of section \ref{SCTPM}). 
Furthermore,  in this case, $H_0$ being integrable,  the SC keeps oscillating even for higher values of $u$.  

\textbf{Case-3.} Finally we consider the case when both $H_0$ and $H_1$  are chaotic, and the evolution of the SC with respect to $u$ for different 
fixed values of $\tau$ are shown in Fig. \ref{fig:chao_chao}.  In this case, we see that the SC shows prominent linear growth 
of small values of $u$ and saturation at large values $u$, and there is a peak in between.\footnote{The peak is not clearly visible from the plots shown in Fig. \ref{fig:chao_chao}, due to the fact that the dip (also called the correlation hole) in the corresponding 
	survival probability is not prominent. This behaviour is similar to the one shown by time evolution of the SC in interacting 
	realistic spin chains \cite{SCint}.} 
These features are consistent with  the behaviour of time evolution of 
the SC for  interacting chaotic systems after sudden quenches \cite{SCint}.  We also notice that the peak in the SC is present irrespective  
of the values of $\tau$, i.e., the initial state. 

\begin{figure}
	\centering
	\includegraphics[width=0.5\textwidth]{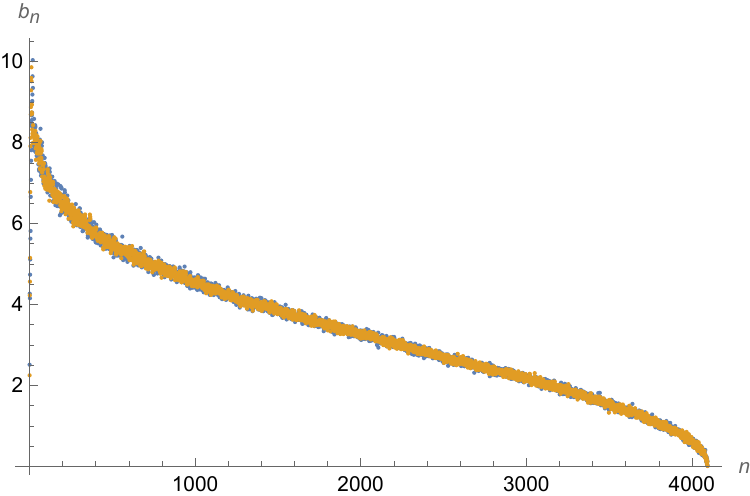}
	\caption{Plot of the second set of LCs ($b_n$) for different values of $\tau$,  when $H_0$ and $H_1$ are both chaotic. The blue points are with
		$\tau=0.1$, and the yellow points are with  $\tau=500$. All other parameters are the same as in Fig. \ref{fig:chao_chao}.}
	\label{fig:bncase3}
\end{figure}

It can also be seen  that, similar to the observation
made in case-2, for  large values of $\tau$, the SC profiles almost merge into each other.  This can once again be explained by noting 
that the initial state $\big|O_0, \tau\big>$, with fixed values of $\tau$, is fixed by the chaotic Hamiltonian $H_1$ in both  the cases. 
Furthermore, the difference between Fig. \ref{fig:int_chao} and Fig. \ref{fig:chao_chao} is also clear from the respective 
plots. When $H_0$ is chaotic, the linear growth of the SC continues up to long times compared to when $H_0$ is integrable, and
in the latter case the SC profile shows oscillations even at late $u$ values, whereas in the first case, the oscillations dry out quickly.
In Fig. \ref{fig:bncase3}, we have also plotted the behaviour of the second set of LCs ($b_n$)
for two different values of $\tau$, i.e., for two different initial states of the $u$-evolution. The pattern for the $b_n$ follows the usual 
behaviour of $b_n$s for a chaotic Hamiltonian \cite{Balasubramanian:2022tpr}.  Furthermore, due to the chaotic nature of the Hamiltonian $H_1$,
the sets of $b_n$s for two different initial states are almost identical (as can also be seen by plotting the histogram of $\log(b_n/b_{n+1})$), 
so that the corresponding SC profiles are also almost identical.

From the discussion of the three cases, we see that only when the Hamiltonian $H_0$ generating  the circuit time evolution is chaotic, the SC saturates 
at large values $u$.  However, the post-quench Hamiltonian $H_1$ also has  important effects on $u$-evolution through the initial state,
and depending on whether it is integrable or chaotic, the SC profile for different $\tau$ can be different. In particular, when $H_1$ is chaotic,
for higher values of $\tau$, the magnitudes of the SC are almost equal, even though they can oscillate or saturate, depending on whether
$H_0$ is integrable or chaotic.

\section{Inverse Participation ratio and the Lanczos coefficients}\label{iprb1}
In this section, we discuss an interesting connection between the IPR and the LCs.
As shown in \cite{Erdmenger:2023shk}, the initial growth of SC is not sensitive to whether the system Hamiltonian is chaotic or integrable. Here
we show that the early evolution of the SC can actually be explained from the behaviour of the IPR  of the state $\big|O_0, \tau\big>$ in the 
eigenstates of the $O$ operator. We compute the IPR and the first few LCs 
when $H_{1}$ is taken as an element of the  $su(1,1)$ algebra.

We consider a Lie algebraic model similar to the one discussed in  section \ref{algebra}; however, here we explicitly specify the form
for the system Hamiltonians to be  $H_0 = K_{0}$ and $H_1 = \alpha (K_{+} + K_{-} ) + K_{0}$, where the $K_i$ are the generators of  $su(1,1)$ algebra, and $\alpha$ is a real constant. 
Furthermore, we assume the perturbation operator
to be $W=\exp \big[i \big(K_+ + K_-\big)\big]$. We first want to compute the time evolved Heisenberg 
picture operator  $W_{\tau}  = e^{- i H_{1} \tau} W e^{ i H_{1} \tau}$. 
This  can be easily done by repeatedly applying $su(1,1)$ decomposition formulas \cite{Ban} to get a decomposition of $W_\tau$ of the form
\begin{equation}\label{decomp}
\begin{split}
	e^{- i H_{1} \tau} W e^{ i H_{1} \tau}  =  \exp(A_{+} K_{+}) &\exp(\ln(A_{0}) K_{0}) \\
	&\times \exp(A_{-} K_{-})~.
\end{split}
\end{equation}
It is possible to write down the exact analytical formulas for the functions $A_i(\tau)$. However, these are extremely complicated, we do not show them here.

We are also interested in finding out the IPR of the state $\big|O_0, \tau\big>$ in terms of the eigenstates of $H_0$. 
For the above choice of the Hamiltonian $H_0$, this is given by the following formula 
\begin{equation}
	\text{IPR} =   \sum_{n=0}^{\infty} \big| \big< h,n      \big|O_0 ,\tau \big>    \big|^{4} ~.
\end{equation}
The IPR is a measure
for the level of delocalisation of an initial state -  a small value of IPR indicates that the initial state  is delocalised in the 
basis $|h,n \big>$. It is used to verify whether the local density of states (which is a quantity very similar to  the  quantity
$P(\mathcal{W})$, the probability distribution of  work done in a quench) of a many-body Hamiltonian is ergodically filled or not \cite{Izrailev}.

Assuming that the state after first measurement is $\big|h,0\big>$ (with $h=1/4$, as before), and  using the decomposition in Eq. \eqref{decomp} above, we have the expression for $\big|O_0 ,\tau \big>$ to be
\begin{equation}
	\begin{split}
		\big|O_0 ,\tau \big>   =  \exp(A_{+} K_{+}) \exp(\ln(A_{0}) K_{0}) \big|h,0\big>~\\
		=(A_{0})^{1/4} \sum_{n=0}^{\infty} \frac{(A_{+})^n}{n !}   (K_+)^n \big|h,0\big>~,
	\end{split}
\end{equation}
so that the IPR is given by the summation 
\begin{equation}
	\text{IPR} =   \sum_{n=0}^{\infty} \big| \mathcal{N}_n (A_{0}(\tau))^{1/4}  (A_{+}( \tau))^n   \big|^{4} ~.
\end{equation}
Using the analytical expressions for the functions $A_i(\tau)$, the IPR can be obtained for different values of the  constant $\alpha$
as function of $\tau$.

\begin{figure}
	\centering
	\centering
	\includegraphics[width=0.5\textwidth]{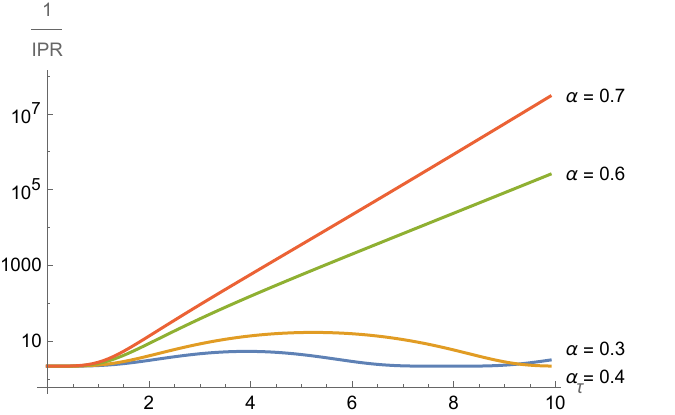}
	\caption{Plot of the participation ratio $\frac{1}{\text{IPR}}$ when $H_1$ is integrable or chaotic for different fixed values of the constant  $\alpha$.}
	\label{fig:ipr}
\end{figure}

Next we compute the LCs in this model by using the ACF
\begin{equation}
	G(u,\tau) = \big< h,0 \big| W_{\tau}^{\dagger} e^{- i H_{0} u}  W_{\tau} \big|h,0\big>~.
\end{equation}
To calculate the ACF, we can  follow a similar approach   as above, and consider  the following decomposition
\begin{equation}
\begin{split}
	W_{\tau}^{\dagger} e^{- i H_{0} u}  W_{\tau} =  \exp(B_{+} K_{+}) &\exp(\ln(B_{0}) K_{0})\\ &\times \exp(B_{-} K_{-})~,
\end{split}
\end{equation}
where, the functions $B_{i}(u, \tau)$ are now functions of both $u$ and $\tau$. Once again, their analytical expressions are 
complicated to provide here.  In terms of these functions, the expression for the ACF simplifies to
\begin{equation}\label{aut}
	\begin{split}
		G(u,\tau) = \big< h,0 \big|  \exp(\ln(B_{0}(u,\tau)) K_{0})\big|h,0\big>~ \\ = B_{0}(u,\tau)^{1/4}.
	\end{split}
\end{equation}
Using this expression for the ACF  we can calculate all the LCs recursively \cite{Balasubramanian:2022tpr}. However, since the expressions for $B_{-}$ , $B_{0}$, and $B_{+}$ 
are complicated, it is only feasible to compute the first few LCs.  These LCs are sufficient for our discussions below.

\begin{figure}
	\centering
	\centering
	\includegraphics[width=0.5\textwidth]{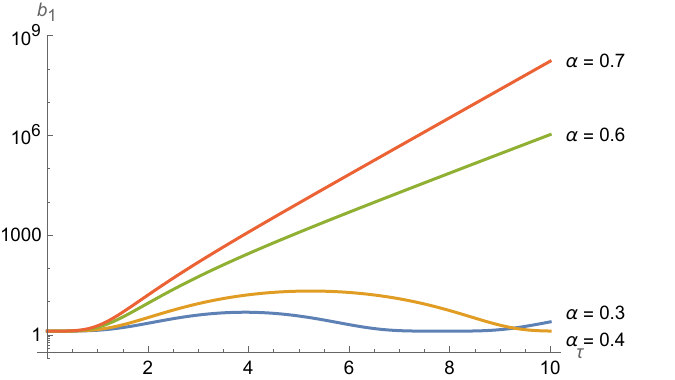}
	\caption{Plot of the first LC $\tilde{b}_{1}$ when $H_1$ is integrable and chaotic for different fixed values of the constant $\alpha$.}
	\label{fig:b1}
\end{figure}

We now compare the behavior of IPR and LCs obtained above as functions of $\tau$  for different values of $\alpha$. In Figs.
\ref{fig:ipr} and \ref{fig:b1} we have plotted
the participation ratio ($\text{PR}=1/\text{IPR}$), and the LC $\tilde{b}_1$ for different fixed values of $\alpha$ below and above 
$0.5$. From these plots 
we observe that $\frac{1}{\text{IPR}}$ and $ \tilde{b}_{1}$ show oscillatory behavior whenever $\alpha$ is less than $0.5$ (indicating 
$H_1$ is stable), whereas,
whenever $\alpha$ is greater than $0.5$, both $\frac{1}{\text{IPR}}$ and $\tilde{b}_{1}$ show exponential growth with $\tau$. This definitive
change in the  behaviour of the LCs and the IPR can be explained mathematically by noticing that, from the decomposition formulas, the quantity that determines the behaviour of the functions $A_i$ and $B_i$ is $\sqrt{\alpha^2-\frac{1}{4}}$ (this quantity actually appears
as the argument of cosine and sine functions). Therefore, when $\alpha$ crosses $0.5$, all the oscillating terms become hyperbolic, resulting 
in the growth of IPR and LCs.   
Physically, this growth is due to the difference in the ability of the $W_{\tau}$ to spread the initial state over all the eigenstates of $H_0$, i.e., coherence generating power, as discussed in \cite{Anand:2020qhf}. This connection shows that in this scheme, the initial growth of the SC is sensitive to the nature of $H_1$.  Here, we have illustrated this connection by considering $H_1$ as an element of a Lie algebra; however, it will be interesting to see whether this conclusion is true for realistic quantum chaotic systems as well.

\section{Summary and Conclusions }\label{conclusion}

In this paper we have extended the concept of complexity of unitary evolution to the cases when an arbitrary initial state in a TPM protocol evolves under a backward and forward evolution through a Hamiltonian, 
with a perturbation applied in-between. Therefore, this quantity measures how an initial eigenstate of a Hermitian
operator $O$ spreads in the Hilbert space corresponding to a hypothetical system whose Hamiltonian is the operator $O$, i.e., it is the complexity of  the $u$-evolution, where the parameter $u$ is the Fourier conjugate to the eigenvalues of the observable $O$, 
also known as the second time of evolution in the literature.
As we have argued here, this definition of the SC of the $u$-evolved state is, in a sense, a bit different from the usual SC of a
	time-evolved state under
	a Hamiltonian evolution, since here, the initial state (or, in a different interpretation, the generator of the circuit evolution) is determined 
	the time-evolved perturbation operator $W$ by a system Hamiltonian $H$. Thus, the nature of this perturbation $W$, as well as the Hamiltonian
	$H$ also play crucial roles in determining the SC of the $u$-evolved state. Hence, as we have verified in the numerical examples in sec. \ref{spinchain},  
the complexity of the $u$-evolved state  incorporates both 
the operator complexity or the KC of the perturbation $W$ (through the initial state), as well as the SC under the $u$-evolution. 
For a sudden quench scenario -  the one we have considered 
in this paper, the Hamiltonian $H$ is the post-quench Hamiltonian $H_1$ of a quantum system, whereas the operator $O$ is the pre-quench Hamiltonian $H_0$.

Our starting point has been an identification made in \cite{Campisi}, where it was shown that the CF of the probability distribution
of the difference in the eigenvalues of an operator $O$ measured in a TPM protocol is the OTOC between the time-evolved (Heisenberg picture)
perturbation operator $W$, and the exponential of the operator $O$ itself. Here, we have first identified this OTOC as the ACF
corresponding to the $u$-evolution. 
Then, as the next natural step, we have studied the spread of an initial state under this 
	dual or second time of evolution (also referred to as the circuit time in this paper to distinguish it from the usual time associated with evolution by the system Hamiltonian). The motivation for this has been the well-known fact that (the complex conjugate of)
	the ACF is the first coefficient among a set of probability amplitudes when a unitaryly evolved state is written in terms of the Krylov basis 
	generated by the corresponding generator of the unitary evolution.
Furthermore, we have shown that, when $O$ is the Hamiltonian $H_0$, one can actually identify the LCs associated with the $u$-evolution 
with certain state-dependent cost functions, such as the FS metric defined on the space of pure state. These relations, on one hand,
 provide clearer interpretations of these coefficients, while on the other hand, show that, it might be possible to directly make 
  connections between  SC and geometric measures of circuit complexity  of an unitarily evolved state.\footnote{In this context, we note that in a very 
 recent work \cite{Aguilar-Gutierrez:2023nyk}, the authors have shown that SC is not a measure of distance. Thus, 
	unlike Nielsen's or FS complexity, the SC of a time-evolved state  can not be understood as a geodesic distance between two points 
	on some  metric space. It will be interesting to find out the implications of this result in view of the above mentioned connection between 
	the LCs and certain cost functions for Nielsen's complexity geometry.}

In the later part of this paper, we have used this definition of the SC to study the evolution of states with respect to the circuit time in both analytical and numerical examples, and illuminated new relationships between different well-known information theoretic quantities,
such as the FOTOC and the quantum Fisher information. The analytical
example we have studied uses the Lie algebraic CS method of obtaining the Krylov basis, LCs as well as the SC. For the $su(1,1)$ and $su(2)$
Lie algebras, we have shown how to obtain the LCs and the Krylov basis corresponding to the unitary evolution generated by the observable 
$O$. 

The next example we have considered is that of a TPM protocol with a sudden quench, where the backward and forward time evolution 
is generated by a post-quench Hamiltonian $H_1$, whose parameters are different from that of the pre-quench Hamiltonian $H_0$. Furthermore, 
$H_0$ is taken as the observable $O$, so that the circuit evolution (or the $u$-evolution) is generated by $H_0$ itself.  
The quantum many-body system we consider 
is the integrable and the chaotic limits of an Ising chain in the presence of a transverse field, and for different values of the parameter 
$\tau$ (which fixes the initial state of the circuit evolution), we numerically obtain the SC of the state generated by $H_0$.
From the results we obtain it can be concluded that, only when $H_0$ is chaotic, the SC saturates at late circuit time. On the other hand,
when $H_1$ is chaotic, it affects the SC of large $\tau$ initial states : namely, for this case, the magnitudes of the 
profile for the SC are identical. 

Before concluding, here we outline some further implications of the results presented in our paper, as well as some important future
directions that can be perused. 
Firstly, the complexity of the spread of an initial state with respect to $u$-evolution, 
that we have introduced here can be thought of an associated with a hypothetical auxiliary
system whose evolution is generated by the observable measured in the TPM protocol (and the role of time is played by the parameter
conjugate to the eigenvalues of the observable). Since, the ACF in such a system is the Fourier transform of the WD, it is natural to ask
what could be the quantity which is conjugate to the SC. Since one can write a Hermitian operator corresponding to SC, one way to approach this problem is to study the statistics of the SC itself and then consider the Fourier transform of that quantity.  For chaotic or 
integrable Hamiltonians, this distribution should  contain important information about the nature of these Hamiltonians \cite{Chenu1, Chenu2}.

Secondly, the statistics of SC, as mentioned above, also points out a possible connection of the results discussed in this 
	paper with experimentally measurable quantities. First,  we notice that if we want to measure the statisics of work  done (or by) a system, 
	we have to keep track of the transitions between energy levels of a quantum system before and after a certain process. This is 
	of course true for measuring statistics of some other observable as well, i.e., we have to analyse the change in the eigenvalues 
	of that observable before and after a process (see the definition of distribution for  a general observable in a TPM scheme, Eq. \eqref{pdfo}).
	Though, for a generic quantum many-body system, keeping track of transitions between energy levels due to a non-equilibrium process (such as a quantum
	quench) is extremely difficult, for relatively simple quantum systems, the probability distribution of the work done has actually been 
	measured in experimental setups, such as trapped atoms and ultracold atoms (see refs. \cite{Batalhao, An, Cerisola}
	for discussions on such experiments). From such measured  WD, one can directly obtain the corresponding CF by using the Fourier transform
	of this distribution (in fact, this was the scheme
	proposed in \cite{Campisi} to measure OTOCs without using any ancillary systems),  as well as LCs (see \cite{Pal:2023yik}).
	Now, since CF is the first of the set of coefficients $\tilde{\phi}_n$, we can, in turn, study Fourier transforms of the higher 
	order $\tilde{\phi}_n$ as well, not only just $\tilde{\phi}_0$, and see whether those Fourier transforms
	have some physical meaning in a TPM protocol 
	and whether these can be measured in experimental setups mentioned above which are used measure the WD. 
	Furthermore, since the SC is just the weighted sum of  the modulus square of  $\tilde{\phi}_n$s, we believe it might provide a way of measuring SC itself
	from such experiments.  Finding out the significance of Fourier transforms  of each $\phi_{n}$ as a  possible distribution of some
	quantum mechanical observable, and understanding their subsequent experimental significance is an interesting problem, and we hope 
	to report on this in the future.

\begin{center}
\bf{Acknowledgements}
\end{center}
The work of TS is supported in part by the USV Chair Professor position at the Indian Institute of Technology, Kanpur.

\appendix

\section{Different interpretations of the OTOC for a TPM protocol}\label{interpretations}
In sections \ref{SCTPM}, \ref{algebra}, and \ref{spinchain}, we interpreted the OTOC in Eq. \eqref{CF-OTOC} as ACF  in two different ways. In particular, 
in the interpretation used in section \ref{SCTPM} to define the SC of the $u$-evolution, as well as in  the numerical computations we did in section \ref{spinchain}, the circuit evolution is generated by the operator $O$ itself, while the initial state it acts on depends on $\tau$ - the time 
after which the perturbation is applied in a TPM protocol, through the operator $W_\tau$. We call this the case-1.
On the other hand,  in section \ref{algebra}, the $u$-evolution is generated by an `effective' observable $O_{\text{ef}}(\tau)$, which
itself is dependent on $\tau$, and is determined  the Hamiltonian $H_1$
as well as the perturbation $W$, while the initial state on which this evolution operator acts is itself independent of $\tau$. This is designated
as case-2 below.  
In this Appendix, we briefly discuss the relationship between the SC in these two cases and show that, in fact, the complexities 
defined in these two ways are equal -  we can use any of them without affecting the physical conclusions.

First we write down the $u$-evolved state in two cases ($W_\tau$ is Heisenberg picture time-evolved  version of the the unitary perturbation 
operator $W$),
\begin{equation}\label{uevolved}
	\begin{split}
			\big|\Psi ^{1}(u)\big>=e^{-iu O}\big|O_0, \tau\big>=e^{-iu O}W_\tau \big|O_0\big>~,~~\text{and}\\
			\big|\Psi^{2}(u)\big>=e^{-iu O_{\text{ef}}}\big|O_0\big>=W_\tau^ \dagger e^{-iu O}W_\tau \big|O_0\big>~.
	\end{split}
\end{equation}
Both of these evolved states can be expanded in terms of the Krylov basis generated by the operators $O$ and $O_{\text{ef}}$ respectively
so that we have\footnote{For convenience, we suppressed the dependence of different quantities on $\tau$ in the following. They can be 
	brought back appropriately from the contexts of cases we will consider.}
\begin{equation}\label{expansion2}
	\big|\Psi ^{i}(u)\big>= \sum_{n}\tilde{\phi}^{i}_{n}(u)|\tilde{K}^i_{n}\rangle~, ~~~i=1,2~.
\end{equation}
Furthermore, comparing the two evolved states in Eq. \eqref{uevolved}, we see that they are related by the unitary transformation 
$\big|\Psi^{2}(u)\big>=W_\tau ^ \dagger(\tau)\big|\Psi ^{1}(u)\big>$.

Next, we notice that, since the ACF is the same in both the cases, the corresponding LCs obtained from the moments of the 
ACF are actually the same (this statement will also be verified later). 
Therefore, considering the action of the operator $O$ on the Krylov basis $|\tilde{K}^1_{n}\rangle$, as in Eq. \eqref{Krylov-basis},
we obtain  the following equality 
\begin{equation}\label{Krylov2}
	W_\tau ^ \dagger \big| \tilde{K}^1_{n+1} \big>=\frac{1}{\tilde{b}_{n+1}}\big[(O_{\text{ef}}-\tilde{a}_{n}) W_\tau ^ \dagger\big|\tilde{K}^1_n\big>-\tilde{b}_{n}W_\tau ^ \dagger\big|\tilde{K}^1_{n-1}\big>\big]~.
\end{equation}
Now comparing this with the action of the operator $O_{\text{ef}}$ on $|\tilde{K}^2_{n}\rangle$, we see that, as expected, two sets of Krylov basis
are related by the transformation $|\tilde{K}^2_{n}\rangle=W_\tau ^ \dagger(\tau) |\tilde{K}^1_{n}\rangle$. Using this relation and the expansion of the 
evolved states in the Krylov basis in Eq. \eqref{expansion2}, it is easy to see that the exasperation coefficients are actually the  same in both cases :
$\tilde{\phi}^{1}_{n}(u)=\tilde{\phi}^{2}_{n}(u)$, so that the corresponding SC, defined as the weighted sum of their modulus squared are also 
equal. This proves the equivalence in terms of the SC of the two interpretations of OTOC  used in the main text.
Furthermore, the fact that the LCs are actually equal in two cases can be verified from the definition of $\tilde{a}_n$s
 (in Eq. \eqref{an}) and 
$\tilde{b}_n$s (which are just the normalisation constants associated with Krylov basis) and noting that, as established above for two operators
related by $O_{\text{ef}}=W_\tau ^ \dagger O W_\tau$, the corresponding Krylov basis vectors are related by $|\tilde{K}^2_{n}\rangle=W_\tau ^ \dagger(\tau) |\tilde{K}^1_{n}\rangle$.


\end{document}